\documentclass[pra,twocolumn,superscriptaddress,showpacs,preprintnumbers,amsmath,amssymb]{revtex4}

\usepackage{graphicx}
\usepackage{dcolumn}
\usepackage{bm}
\usepackage{enumerate}

\begin{document}

\bibliographystyle{apsrev} 

\title{Periodically modulated electromagnetically induced transparency}

\author{Yingying Han}
\affiliation{School of Physics and Technology, Wuhan University, Wuhan, Hubei 430072, China}

\author{Jun Zhang}
\affiliation{School of Physics and Technology, Wuhan University, Wuhan, Hubei 430072, China}


\author{Wenxian Zhang}
\email[Corresponding email: ]{wxzhang@whu.edu.cn}
\affiliation{School of Physics and Technology, Wuhan University, Wuhan, Hubei 430072, China}

\date{\today}

\begin{abstract}
Phenomena of electromagnetically induced transparency (PEIT) may be interpreted by the Autler-Townes Splitting (ATS), where the coupled states are split by the coupling laser field, or by the quantum destructive interference (QDI), where the atomic phases caused by the coupling laser and the probe laser field cancel. We propose modulated experiments to explore the PEIT in an alternative way by periodically modulating the coupling and the probe fields in a $\Lambda$-type three-level system. Our analytical and numerical results rule out the ATS interpretation and show that the QDI interpretation is more appropriate for the modulated experiments. The proposed experiments are readily implemented in atomic gases, artificial atoms in superconducting quantum devices, or three-level meta-atoms in meta-materials.
\end{abstract}

\pacs{42.50.Gy, 42.50.Nn, 42.50.Md}
\maketitle

\section{Introduction}

Phenomena of electromagnetically induced transparency exists in a wide variety of physical systems, such as atomic gases~\cite{PhysRevLett.66.2593,ISI:000325006700104,PhysRevLett.74.666}, artificial atoms in superconducting quantum circuits~\cite{PhysRevLett.104.163601}, quantum dots~\cite{ISI:000259686400011}, optomechanics~\cite{ISI:000289199400038}, and three-level meta-atoms in meta-materials~\cite{PhysRevLett.101.047401,PhysRevLett.101.253903}. Important applications of EIT include the slow light experiments~\cite{PhysRevLett.86.783,ISI:000303597400005,PhysRevLett.64.1107,PhysRevA.75.063818}, quantum memory~\cite{PhysRevA.80.014301,PhysRevA.91.053805}, precision measurements~\cite{PhysRevLett.95.087001,PhysRev.124.1866}. However, the theoretical interpretations of the PEIT has not been unified~\cite{doi:10.1080/09500349808231909,RevModPhys.77.633}. Among these many explanations, two theories, the ATS~\cite{PhysRev.100.703, ALB} and the QDI~\cite{ISI:000357418300009}, are often quoted.

According to the ATS theory, the strong coupling field causes a large splitting between the doublet structure in the absorption profile and the probe field is unabsorbed in the space within the doublet~\cite{0022-3700-10-12-010,PhysRevA.87.033835}. Alternatively, the QDI theory considers that the quantum destructive interference of two or many transition paths results in the atomic transparency~\cite{:/content/aip/magazine/physicstoday/article/50/7/10.1063/1.881806}. The difference between the ATS and the QDI theory has also been investigated~\cite{PhysRevLett.107.163604,PhysRevA.87.013823,PhysRevA.81.053836}. The common conclusion is that the ATS (QDI) dominates if the coupling field is strong (weak), compared to the decay of the three-level system. A crossover exists in between~\cite{PhysRevA.87.043813, PhysRevA.89.063822}, where both theories do not work well. Therefore, new experiments are demanded in order to unambiguously discern the two theories.

By periodically switching on and off the coupling and/or the probe field, we may distinguish the two theories in a different manner. In fact, switching on suddenly the
coupling field in a $\Lambda$-type three-level system shows unusual transient gain features~\cite{PhysRevA.65.053802}. Moreover, modulated two-level systems often exhibit quite different dynamics, compared with the free evolution~\cite{PhysRevLett.98.013601}. For example, the decoherence of a two-level qubit is strongly suppressed by periodically rotating the qubit~\cite{PhysRevLett.114.190502, PhysRevB.75.201302, PhysRevB.77.125336}. We thus expect different dynamics of the three-level system by modulating the coupling and/or the probe field.

A key feature of the modulation in the three-level system is that the ATS disappears if the coupling field is off, as shown in Fig.~\ref{fig:pe}. The probe field is then absorbed and PEIT should disappear according to the ATS theory (see also Table.~\ref{tab:atsqdi}). However, the phases induced by the coupling field and the probe field may cancel even if the two fields are not simultaneously on. The PEIT might occur according to the QDI theory.

In this paper, we put forward two modulated EIT experiments. We investigate the absorption of a $\Lambda$-type three-level system~\cite{PhysRevA.51.4959} driven by periodically modulated coupling and probe fields. By comparing the analytical and numerical results with the ATS and QDI's predictions, we expect to distinguish these two theories in the modulated EIT experiments. We also carry out numerical calculations of the modulated EIT under real experimental conditions, i.e., nonzero decay and mixed initial state, and show that it is practical to realize the proposed modulated EIT experiments with current techniques.

The paper is organized as follows. In Sec.~\ref{sec:sys}, we describe the proposed experiment by modulating the coupling and the probe laser firld in a $\Lambda$-type three-level atom. The main analytical and numerical results for various modulation situations are presented in Sec.~\ref{sec:rd}. We also discuss the validity of the interpretations of the ATS and the QDI for the proposed experiments. We draw our conclusion in Sec.~\ref{sec:con}.

\section{Modulated EIT in a $\Lambda$-type three-level system}
\label{sec:sys}

\begin{figure}
\includegraphics[width=3.2in]{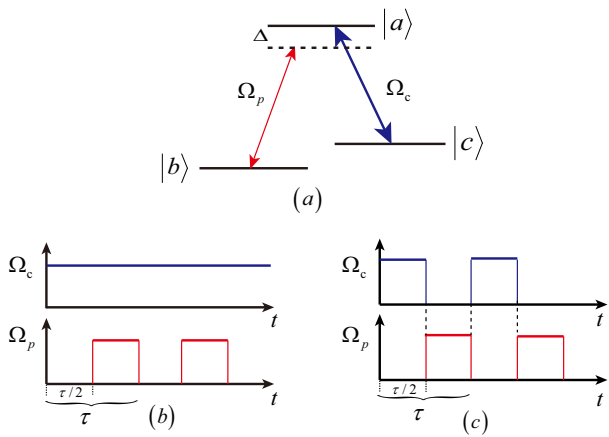}
\caption{\label{fig:pe} (Color online) Schematic of modulated EIT in a $\Lambda$-type three-level atom. (a) Levels of the atom, the resonant coupling field $\Omega_{c}$, and the probe field $\Omega_{p}$ with a detuning $\Delta$. (b) Pulse sequence of the lasers for single-modulation, where only the probe field is modulated, and (c) double-modulation, where both the coupling and the probe fields are modulated complementarily.}
\end{figure}

We consider a $\Lambda$-type three-level atom shown in Fig.~\ref{fig:pe}(a). A strong laser resonantly couples the ground state $|c\rangle$ and the excited state $|a\rangle$ with a Rabi frequency $\Omega_c$. A second laser couples the state $|b\rangle$ and $|a\rangle$ with a Rabi frequency $\Omega_p$ and a detuning $\Delta$. The transition between states $|b\rangle$ and $|c\rangle$ is forbidden.

For a standard EIT system, the coupling and the probe lasers are on simultaneously~\cite{PhysRevLett.66.2593,PhysRevLett.64.1107}. The Hamiltonian of such a system is
\begin{eqnarray*}
 H_{EIT} &=& H_0 -\frac{1}{2}(\Omega_{p}|a\rangle\langle b|
           +\Omega_{c}|a\rangle\langle c|+h.c.)
\end{eqnarray*}
where $H_0 = (\Delta/2)(|a\rangle\langle a|-|b\rangle\langle b| +|c\rangle\langle c|)$. We have set $\hbar = 1$ and adopted the rotating wave approximation~\cite{QO}. Since the initial state is a dark state at $\Delta = 0$,
\begin{equation}
\label{eq:ds}
|\Psi(0)\rangle=\frac{\Omega_{c}|b\rangle-\Omega_{p}|c\rangle}{\sqrt{\Omega_{p}^2+\Omega_{c}^2}},
\end{equation}
the atomic gases are transparent for the probe laser due to the existence of the coupling laser.

We focus on two modulation situations: (I) single-modulation where only the probe field is switched on and off periodically and (II) double-modulation where both the coupling and the probe fields are switched complementarily. In the single-modulation situation as shown in Fig.~\ref{fig:pe}(b), the time-dependent system Hamiltonian is
\begin{equation}\label{eq:hI}
 H_{I}=\left\{ \begin{array}{ll}
 H_{1}, \;& \textrm{$t\in[n\tau,(n+\frac{1}{2})\tau]$}\\\\
 H_{2}, \;& \textrm{$t\in[(n+\frac{1}{2})\tau,(n+1)\tau]$}
   \end{array} \right.
\end{equation}
where
\begin{eqnarray*}
  H_1 &=& H_0 -\frac{1}{2}\left(\Omega_{c}|a\rangle\langle c|
            +h.c.\right), \\
  H_2 &=& H_0 -\frac{1}{2}(\Omega_{p}|a\rangle\langle b|
           +\Omega_{c}|a\rangle\langle c|+h.c.)
\end{eqnarray*}
with $\tau$ denoting the cycle period and $n=0,1,2,\cdots$. In the double-modulation situation as shown in Fig.~\ref{fig:pe}(c), the system Hamiltonian is similar to the single-modulation one,
\begin{equation}\label{eq:hII}
 H_{II}=\left\{ \begin{array}{ll}
 H_{1},\; & \textrm{$t\in[n\tau,(n+\frac{1}{2})\tau]$}\\\\
 H_{2},\; & \textrm{$t\in[(n+\frac{1}{2})\tau,(n+1)\tau]$}
   \end{array} \right.
 \end{equation}
 where
 \begin{eqnarray*}
  H_1 &=& H_0 -\frac{1}{2}(\Omega_{c}|a\rangle\langle c|
            +h.c),\\
  H_2 &=& H_0 -\frac{1}{2}(\Omega_{p}|a\rangle\langle b|
            +h.c).\\
\end{eqnarray*}

The modulated EIT experiments proposed here are actually readily realized in many three-level systems, such as atomic gases~\cite{ISI:000303597400005,ISI:000325006700104}, artificial atoms in superconducting quantum circuits ~\cite{ISI:000346274500001, PhysRevA.89.063822}, three-level meta-atoms in meta-materials~\cite{PhysRevLett.107.043901, PhysRevLett.101.253903, PhysRevE.83.046604}. For atomic gases, the Rabi frequency for an atom driven by a strong laser can reach $10^9$Hz. The detuning $\Delta$ is adjustable and ranges from $10$Hz to $10^9$Hz. The pulse period $\tau$ relates to the laser repetition rate and the smallest $\tau$ is about $10$ns~\cite{ISI:000182372200013}. For such systems, the limits $\Delta \,\tau \ll 1$ and $\Omega_{c,p} \,\tau \ll 1$ are feasible.

\section{Results and discussions}
\label{sec:rd}

For a standard EIT system, both the ATS theory and the QDI theory can interpret the transparency observed in atomic gases~\cite{PhysRev.100.703,doi:10.1080/09500349808231909,PhysRevLett.66.2593}. While in the two modulated EIT systems, as shown in Table~\ref{tab:atsqdi} the ATS induced by the coupling field occurs in the single-modulation situation but is absent in the double-modulation situation when the probe field is on. According to the ATS theory, the three-level system would only exhibit PEIT in the single-modulation situation. On the contrary, the QDI exists and the PEIT  appears only in the double-modulation situation. By observing the absorption of the atom in the two modulation situations, we are able to clearly verify the validity of the ATS and the QDI theories.

\begin{table}
\centering
\caption{\label{tab:atsqdi} Predictions from the ATS and QDI theories for the $\Lambda$-type three-level system in the two modulation situations. Our analytical and numerical result confirm the QDI predictions.}
\begin{tabular}{c|c c}
\hline
\hline
 & Single-mod. & Double-mod. \\
\hline
ATS & PEIT & No PEIT\\
QDI &No PEIT & PEIT \\
\hline
\hline
\end{tabular}
\end{table}

The initial state in all situations is set as a dark state, Eq.~(\ref{eq:ds}). We will calculate the fidelity of the system defined as
\begin{equation}\label{eq:fid}
  F(t) = |\langle \Psi(0) | \Psi(t) \rangle | ^2
\end{equation}
and the absorption between the states $|b\rangle$ and $|a\rangle$, which is proportional to the imaginary part of the off-diagonal element of the density matrix $\rho(t)$
\begin{equation}\label{eq:ab}
  {\rm Im} (\chi) \propto {\rm Im}(\rho_{ab}).
\end{equation}

\subsection{Situation I: Single-modulation}

In the single-modulation situation, the time evolution operator for a period is
\begin{equation}
 U(\tau)=e^{-i \tau H_{2} /{2}}e^{-i \tau H_{1} /{2}} = e^{-i\tau H_{\rm eff}}.
\end{equation}
By employing the Baker-Campbell-Hausdorff formula~\cite{ISI:000354870900011} we obtain
\begin{equation}
H_{\rm eff} \approx \frac{1}{2} (H_{1}+H_{2}) - \frac{i\tau}{8}[H_2,H_1] - \frac{\tau^2}{96}[H_2-H_1,[H_2,H_1]].
\end{equation}
We have neglected higher order terms in the limit of small $\tau$. By diagonalizing the effective Hamiltonian $H_{\rm eff}$, we find the total evolution operator at time $t = n\tau$, $[U(\tau)]^n$, and thus the fidelity
\begin{equation}\label{eq:smf}
F(t) = \frac{1}{4A^2} \left|(576-\tau^2)^2 + (36864+768\tau^2+\tau^4) \cos(\frac{\sqrt{A}\,t}{768}) \right|^2
\end{equation}
where $A=184320-192\tau^2+\tau^4$. We have set hereafter $\Omega_c = \Omega_p = 1$ and $\Delta = 0$ unless stated otherwise. As $\tau$ approaches 0, the Eq.~(\ref{eq:smf}) is further simplified as
\begin{equation}\label{eq:ssmf}
  F(t) \approx \left[\frac 9 {10} + \frac 1 {10} \cos(\frac{\sqrt 5} 4 \, t)\right]^2.
\end{equation}
Besides the analytical results Eq.~(\ref{eq:smf}), we also calculate the fidelity numerically by directly integrating the time-depend Hamiltonian Eq.~(\ref{eq:hI}).

\begin{figure}
\includegraphics[width=3.2in]{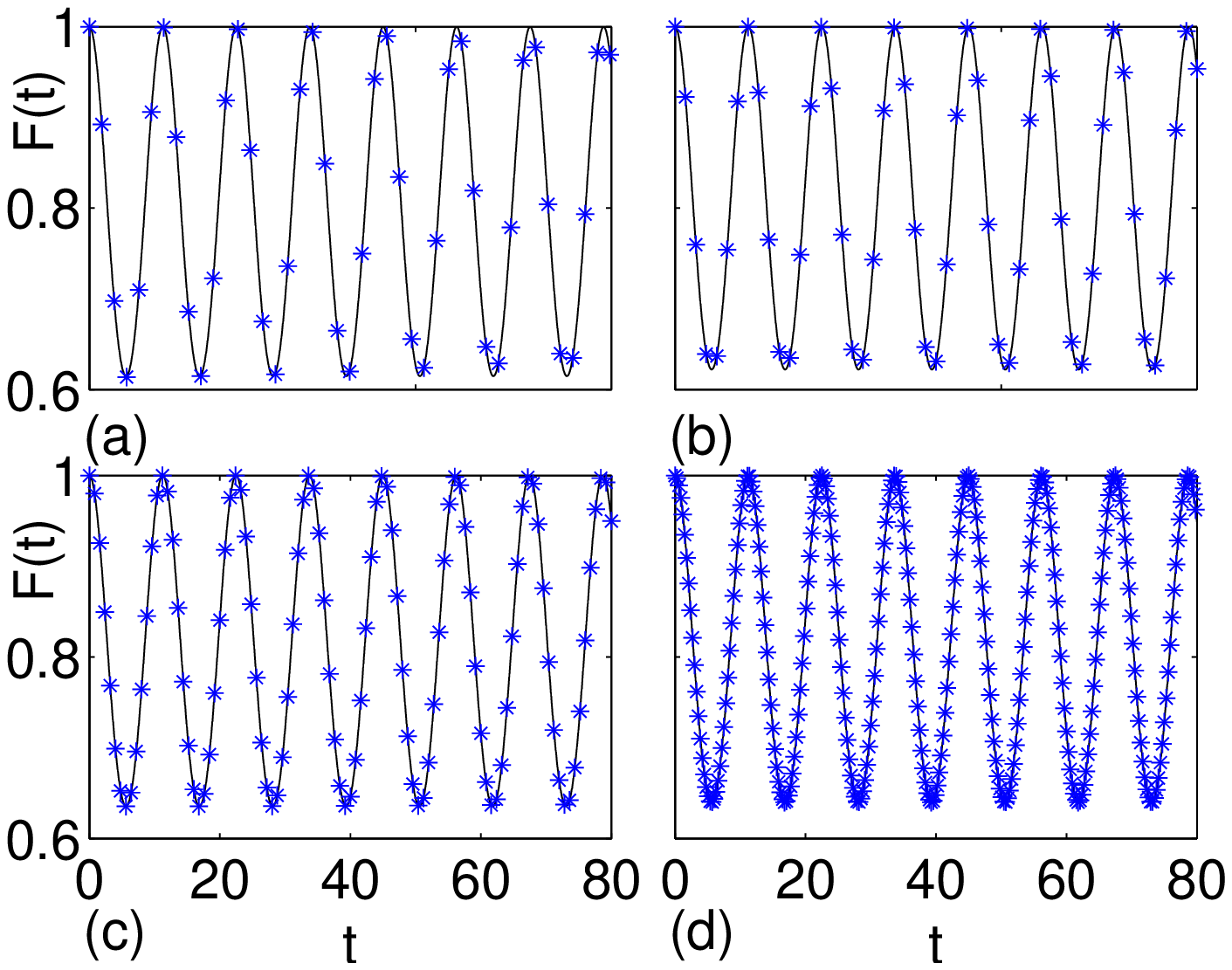}
\includegraphics[width=3.2in]{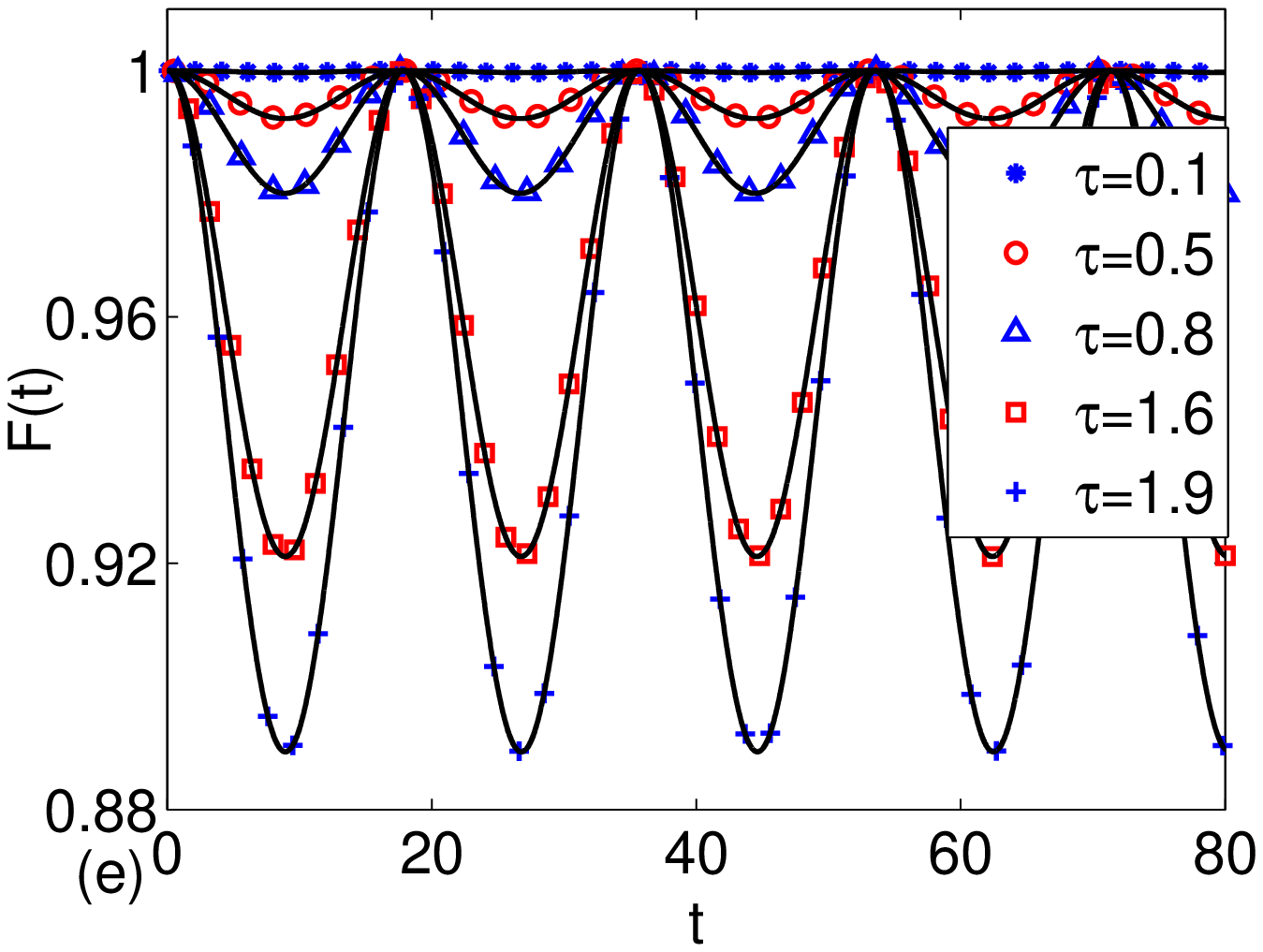}
\caption{\label{fig:fid} (Color online) Time dependence of the fidelity for the single-modulation situation at $\tau = 1.9$ (a), $1.6$ (b), $0.8$ (c), and $0.1$ (d) and for the double-modulation situation (e) at $\tau=0.1$ (asterisks), $0.5$ (circles), $0.8$ (triangles), $1.6$ (squares), and $1.9$ (pluses). The lines in all panels are corresponding analytical predictions from Eq.~(\ref{eq:smf}) and Eq.~(\ref{eq:dmf}). The markers denote the numerical results. Clearly PEIT appears only in the double-modulation situation as $\tau$ approaches 0.}
\end{figure}

\begin{figure}
\includegraphics[width=3.2in]{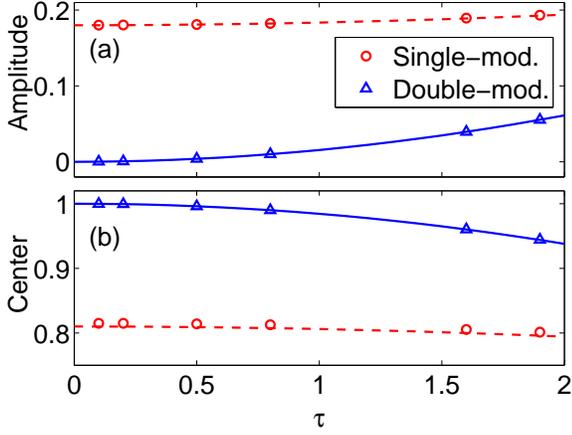}
\caption{\label{fig:aa} (Color online) (a) Dependence of the amplitude and (b) the center of the oscillations of $F(t)$ on the modulation period $\tau$ for two modulations. The markers denote the numerical results and the lines the analytical prediction from Eqs.~(\ref{eq:smf}) and (\ref{eq:dmf}), correspondingly. The average of the double-modulation situation approaches 1 and exhibits PEIT as $\tau \rightarrow 0$.}
\end{figure}

We present the numerical results and analytical predictions on the fidelity $F(t)$ in Fig.~\ref{fig:fid}(a)-(d) for various $\tau$'s. We find from the figure that $F(t)$ oscillates with time and these oscillations are almost independent of $\tau$ [see also Eq.~(\ref{eq:ssmf})]. The numerical results agree well with the analytical predictions from Eq.~(\ref{eq:smf}). We plot the amplitude and the center of these oscillations in Fig.~\ref{fig:aa}. Clearly, the amplitude is not zero and the center is away from 1. These deviations imply that there is no PEIT in the single-modulation situation. Since ATS induced by the coupling field indeed exists in the singe-modulation situation, we are able to rule out the ATS interpretation in the single modulation experiment.

According to the QDI theory, the accumulated phases do not cancel exactly during each period for one path $|b\rangle \rightarrow |a\rangle$ and the other $|c\rangle \rightarrow |a\rangle$. Thus there is no PEIT in the single-modulation situation. Our results support the QDI interpretation for
single modulation experiment as $\tau$ approaches 0.

\subsection{Situation II: Double-modulation}

Next we consider the double-modulation situation where the coupling and the probe fields are on and off complementarily. Since the ATS, induced by the coupling field, and the probe field do not exist simultaneously, the PEIT would never occur according to ATS theory. While according to the QDI theory, the PEIT would occur if the accumulated phases during the first and the second half period cancel exactly.

Similar to the single-modulation situation, we straightforwardly calculate the fidelity for the double-modulation situation and obtain the following analytical result
\begin{equation}\label{eq:dmf}
  F(t) = \frac{1}{4B^2}\left|(192-\tau^2)^2+288\tau^2\cos(\frac{\sqrt{B}\,t}{384})\right|^2
\end{equation}
where $B=18432-48\tau^2+\tau^4/2$. It is easy to check that $F(t)$ approaches 1 if $\tau \rightarrow 0$, indicating the occurrence of the PEIT.

We plot the prediction from Eq.~(\ref{eq:dmf}) in Fig.~\ref{fig:fm}(e) and compare with the numerical results obtained by integrating the time-dependent Hamiltonian Eq.~(\ref{eq:hII}). From this figure we see oscillations again, but the center of the oscillation approaches 1 as $\tau$ decreases and the oscillation amplitude becomes smaller. Such a trend is also shown in Fig.~\ref{fig:aa}, indicating that the PEIT occurs as $\tau \rightarrow 0$.

In the double-modulation situation, our results show that the PEIT appears without the ATS. Once again the ATS interpretation gives an invalid prediction. However, according to the QDI theory, the accumulated phases for the state $|a\rangle$ during the first and the second half period cancel exactly and the PEIT should occur. With the parameters $\Omega_c=\Omega_p = 1$ and the initial state $(|b\rangle -|c\rangle)/\sqrt 2$, the wave function amplitude of the state $|a\rangle$ after the first half period is approximately $(i/\sqrt 2)\sin(\pi+\tau \Omega_c/4)$ and after the second half period $(i/\sqrt 2)[\sin(\pi+\tau \Omega_c/4)\cos(\tau \Omega_p/4)+\sin(\tau \Omega_p/4)]\approx0$. Here we have separated the three-level system approximately into two two-level systems $|a\rangle$ and $|c\rangle$ (or $|b\rangle$), which is obviously valid in the limit $\tau \rightarrow 0$.

\subsection{Off-resonance}

The detuning is usually nonzero in the EIT experiments, $\Delta \neq 0$~\cite{PhysRevLett.66.2593,PhysRevLett.82.5229}. We now consider the double-modulation situation with nonzero but small detuning.

To calculate the fidelity $F(t)$ and the absorption ${\rm Im}(\rho_{ab})$, we introduce two approximations. The first approximation is adopted in the effective Hamiltonian Eq.~(\ref{eq:hII}) where the higher order terms than $O(\tau^2)$ are neglected. In this way, the evolution operator at $t=n\tau$ becomes
\begin{equation}\label{eq:u}
  U(t=n\tau) = V e^{-in\tau D} \,V^{-1},
\end{equation}
where $D$ and $V$ are the eigenvalues and the eigenvectors of the Hamiltonian, $H_{eff} = V D V^{-1}$. The second approximation is employed to expand $V$ and $D$ and to neglect higher order terms than $O(\tau^2)$ or $O(\Delta^3)$. We then obtain the fidelity
\begin{equation}\label{eq:fd}
    F(t)=a_{1}\cos(f_{1}t)+a_{2}\cos(f_{2}t)+a_{3}\cos(f_{3}t)+a_{4}
\end{equation}
where
\begin{widetext}
\begin{eqnarray*}
  a_{1}&=&2\Delta^2+\frac{\tau^2}{128}(1-8\Delta^2)-{\rm sgn}(\Delta)\left[3\sqrt{2}\Delta^3-\frac{\sqrt{2}\Delta\tau^2}{3072}(12+197\Delta^2)\right]\\
  a_{2}&=&2\Delta^2+\frac{\tau^2}{128}(1-8\Delta^2)+{\rm sgn}(\Delta)\left[3\sqrt{2}\Delta^3-\frac{\sqrt{2}\Delta\tau^2}{3072}(12+197\Delta^2)\right]\\
  a_{3}&=&\frac{\Delta^2\tau^2}{64}\\
  a_{4}&=&1-4\Delta^2-\frac{\tau^2}{64}(1-7\Delta^2)
\end{eqnarray*}
and
\begin{eqnarray}\label{eq:fr}
  f_{1}&=&\frac{\sqrt{2}}{4}+\frac{5\sqrt{2}\Delta^2}{16}-\frac{\sqrt{2}\tau^2}{12288}(4+7\Delta^2)+{\rm sgn}(\Delta)\left[\frac{\Delta}{4}-\frac{3\Delta^3}{2}-\frac{\Delta\tau^2}{512}(1-4\Delta^2)\right] \nonumber \\
  f_{2}&=&\frac{\sqrt{2}}{4}+\frac{5\sqrt{2}\Delta^2}{16}-\frac{\sqrt{2}\tau^2}{12288}(4+7\Delta^2)-{\rm sgn}(\Delta)\left[\frac{\Delta}{4}-\frac{3\Delta^3}{2}-\frac{\Delta\tau^2}{512}(1-4\Delta^2)\right] \nonumber \\
  f_{3}&=&\frac{\sqrt{2}}{2}+\frac{5\sqrt{2}\Delta^2}{8}-\frac{\sqrt{2}\tau^2}{6144}(4+7\Delta^2)
\end{eqnarray}
with ${\rm sgn}(\cdot)$ being the sign function.

The absorption is calculated similarly,
\begin{equation}\label{eq:rho}
{\rm Im}(\rho_{ab})=b_{1}\cos(f_{1}t)+b_{2}\cos(f_{2}t)+b_{3}\cos(f_{3}t)
+b_{4}\sin(f_{1}t)-b_{5}\sin(f_{2}t)-b_{6}\sin(f_{3}t)+b_{7}
\end{equation}
where
\begin{eqnarray*}
b_{1}&=&\frac{-\tau}{32}(1-12\Delta^2)-{\rm sgn}(\Delta)\frac{\sqrt{2}\Delta\tau}{256}(4+35\Delta^2)\\
b_{2}&=&\frac{-\tau}{32}(1-12\Delta^2)+{\rm sgn}(\Delta)\frac{\sqrt{2}\Delta\tau}{256}(4+35\Delta^2)\\
b_{3}&=&-\frac{3\Delta^2\tau}{16}\\
b_{4}&=&\frac{\Delta}{2}(1-2\Delta^2)-\frac{\Delta^3\tau^2}{384}
  -{\rm sgn}(\Delta)\left[\frac{3\sqrt{2}\Delta^2}{4}+\frac{\sqrt{2}\tau^2}{6144}(12-109\Delta^2)\right]\\
b_{5}&=&\frac{\Delta}{2}(1-2\Delta^2)-\frac{\Delta^3\tau^2}{384}
  +{\rm sgn}(\Delta)\left[\frac{3\sqrt{2}\Delta^2}{4}+\frac{\sqrt{2}\tau^2}{6144}(12-109\Delta^2)\right]\\
b_{6}&=&\frac{\sqrt{2}\Delta^2}{2}+\frac{\sqrt{2}\tau^2}{2048}(4-33\Delta^2)\\
b_{7}&=&\frac{\tau}{16}(1-9\Delta^2)
\end{eqnarray*}
The frequencies $f_{1,2,3}$ are given in Eq.~(\ref{eq:fr}).
\end{widetext}

\begin{figure}
\includegraphics[width=3in]{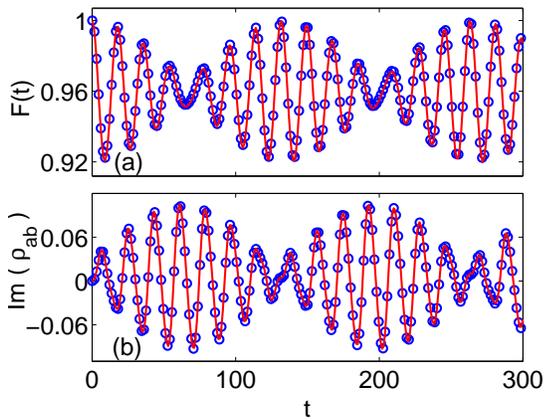}
\caption{\label{fig:te} (Color online) (a) Typical time evolutions of the fidelity $F(t)$ and (b) the absorption ${\rm Im}(\rho_{ab})$ for $\tau=0.1$, $\Omega_{p}=\Omega_{c}=1$, and $\Delta=-0.1$. The markers denote the numerical results and the lines are the analytical results calculated from Eqs.~(\ref{eq:fd}) and (\ref{eq:rho}), correspondingly. The analytical and numerical results agree well.}
\end{figure}

\begin{figure}
\includegraphics[width=3.2in]{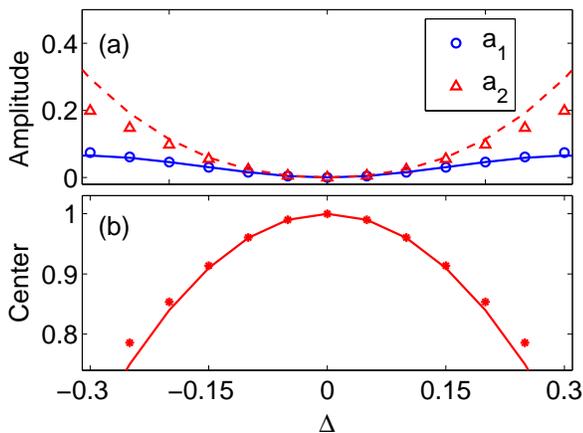}
\caption{\label{fig:fm} Dependence of the (a) amplitudes and (b) center of the oscillations of the fidelity $F$ on $\Delta$ at $\tau=0.1$. The amplitude $a_3$ is much smaller and not shown. The lines are the analytical predictions from Eq.~(\ref{eq:fd}) and the markers denote the numerical results. An EIT-like widow appears in the region of small $\Delta$ where the fidelity is close to 1.}
\end{figure}

\begin{figure}
\includegraphics[width=3.2in]{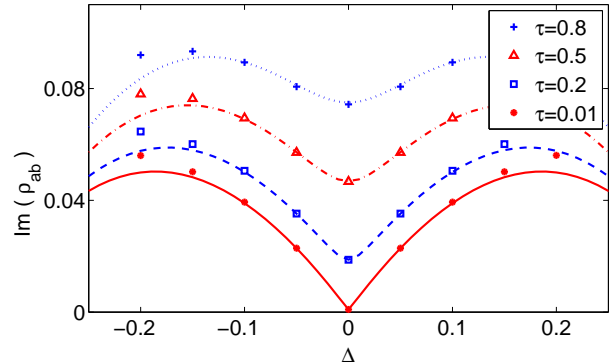}
\caption{\label{fig:am}(Color online) Dependence of the absorption on the detuning $\Delta$ at various $\tau$'s. Clearly, the EIT-like window would be observed in experiments at small $\tau$'s.}
\end{figure}

We plot in Fig.~\ref{fig:te} the analytical as well as numerical results of the fidelity $F(t)$ and the absorption ${\rm Im}(\rho_{ab})$. The analytical predictions agree well with the numerical results, indicating the validity of the adopted approximations. We see clear oscillations and even beats in both $F(t)$ and ${\rm Im}(\rho_{ab}(t))$, which are manifested by that $f_1\approx f_2, a_1\approx a_2$, and $a_3 \approx 0$ for the parameters we choose. Besides the oscillations, the fidelity is close to 1 and the absorption is close to zero, indicating that an PEIT occurs. We also find in the figure that the profile of $F(t)$ and ${\rm Im}(\rho_{ab}(t))$ are negatively correlated, i.e., the higher the $F(t)$ is, the lower the ${\rm Im}(\rho_{ab}(t))$ is, and vice versa.

We extract the oscillation amplitudes $a_1$, $a_2$, and the center $a_4$ from the numerical simulations and plot together with analytical predictions in Fig.~\ref{fig:fm}. At small $|\Delta|$'s, the center of the fidelity is close to 1 and the amplitudes are small, implying an EIT-like widow appears in this region as shown in the figure. However, at large values $|\Delta|\gtrsim 0.2$, the fidelity drops rapidly and the analytical predictions deviate apparently from the numerical results.

We plot the value $b_7 + \sqrt{b_1^2+b_4^2}$ in Fig.~\ref{fig:am}. This value is the main contribution of the absorption and is the observable in a real experiment with population decay (see also Sec.~\ref{sec:dec}). From Fig.~\ref{fig:am}, a perfect PEIT (zero absorption) is exhibited at $\Delta = 0$ and $\tau \rightarrow 0$. As $\tau$ increases, the minimal absorption at $\Delta = 0$ increases but the width of the EIT-like window changes very little. Interestingly, the modulated EIT we discuss here shows similar behavior to the standard EIT.

\subsection{With population decay}
\label{sec:dec}

\begin{figure}
\includegraphics[width=3in]{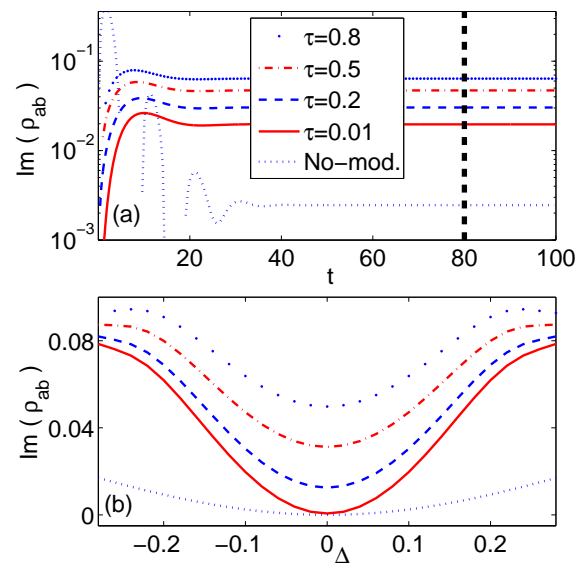}
\caption{\label{fig:td} (Color online) (a) Time evolution of the absorption for various $\tau$'s. Other parameters are $\Omega_{p}=1,\Omega_{c}=1, \gamma=1$, and $\Delta=-0.1$. A plateau appears after long time evolution in the modulated EIT. The vertical dashed line at $t=80$  marks the time to extract the value of the plateau. (b) The absorption ${\rm Im}(\rho_{ab})$ as a function of the detuning $\Delta$ at $\tau = 0.8, 0.5, 0.2, 0.01$ and no-modulation situation from top to bottom at the same time ($t=80$). Similar to the standard EIT curves and Fig.~\ref{fig:am}, an EIT-like window appears.}
\end{figure}

We previously assume the decay in the three-level atoms is zero, but in all kinds of EIT and EIT-like experiments, the decay always exists and plays an important role in some experiments such as atomic EIT~\cite{PhysRevA.80.041805,ISI:000259270600066}. To be practical, we include the population decay in our modulated EIT system and discuss its effect on the observable ${\rm Im}(\rho_{ab})$.

By including the decay process, the problem becomes too complicated to be solved analytically. We thus resort to numerical solution to the following master equation~\cite{QO}
\begin{equation}\label{eq:me}
  \frac{\partial \rho (t)}{\partial t} = i[\rho,H_{II}] -\frac 1 2 \{\Gamma, \rho\}
\end{equation}
with the decay matrix $\Gamma = \gamma |a\rangle \langle a|$ and the anti-commutator operation $\{A,B\} = AB + BA$. We have neglected other minor decay channels without loss of generality. To conserve the total probability, we normalize the system after each modulation cycle. An alternative numerical calculations employing the Lindblad form are described in Appendix~\ref{app:lme}, which agrees qualitatively with the results from Eq.~ (\ref{eq:me}).

Typical time evolutions of modulated system with decay are presented in Fig.~\ref{fig:td}(a). We also plot the results for the standard EIT as a comparison. From panel (a), plateaus appear at long times for both the standard EIT and the modulated ones with different $\tau$. Such plateaus indicate that the three-level system reaches a steady state which is "EIT-like". As the modulation period decreases, the value of the plateau becomes closer to the value for the standard EIT. We also observe oscillations at short times, which is in fact the virtual effect of the sudden turn-on of the probe field~\cite{QO}.

In Fig.~\ref{fig:td}(b) we summarize the dependence of the plateau's value from panel (a) on the detuning. We find obvious EIT-like window for all the small modulation periods. As $\tau$ gets smaller, the EIT-like window becomes more profound with smaller absorption. However, the sizes of the EIT-like window are almost independent of the modulation period. It is quite interesting that the results shown in Fig.~\ref{fig:am} are similar to that in Fig.~\ref{fig:td}(b), not only in the shapes but also in the values. Such coincidence might imply that the decay process is unimportant for the decay range we consider.

\subsection{A mixed initial state}
\label{sec:mix}
\begin{figure}
\includegraphics[width=3in]{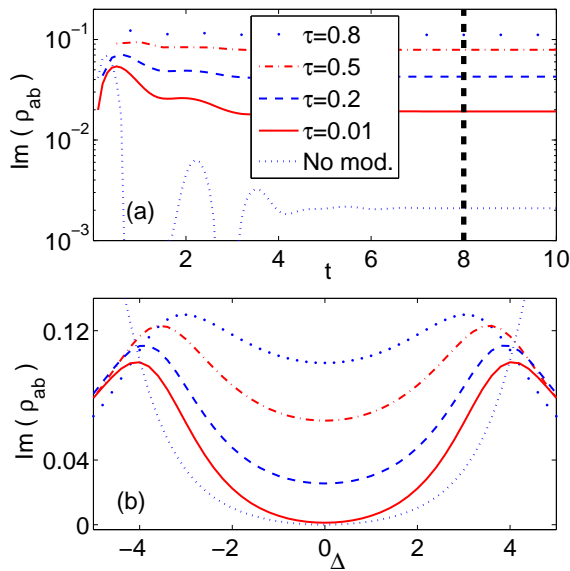}
\caption{\label{fig:meit} (Color online) Same as Fig.~\ref{fig:td} except the initial state being a mixed state with $\rho_{bb}=0.99,\rho_{cc}=0.01$. Other parameters are $\Omega_{p}=1,\Omega_{c}=\sqrt{99}, \gamma=5$, and $\Delta=1$ in panel (a). The vertical dashed line at $t=8$  marks the time to extract the value of the plateau. Similarly, as shown in panel (b), the EIT window would be observed in experiments at small $\tau's$.}
\end{figure}
We previously discuss the PEIT with the initial state being a dark state, which is not the same as in the standard EIT experiment where most atoms are in the ground state $|b\rangle$. To be more practical, we adopt an initial mixed state and set the initial density matrix as $\rho_{bb}=0.99,\rho_{cc}=0.01$, and other elements being zero. Correspondingly, we use $\Omega_{p}=1, \Omega_{c}=\sqrt{99}$, and $\gamma=5$. We calculate the absorption, as shown in Fig.~\ref{fig:meit}, following the same procedure as described in Sec.~\ref{sec:dec}. From Fig.~\ref{fig:meit} we find similar behaviors to Fig.~\ref{fig:td}, indicating that the double-modulated EIT experiment with a mixed state also exhibits PEIT.

\section{Conclusion}
\label{sec:con}

We investigate two situations of modulated EIT by periodically modulating the coupling and the probe fields in a $\Lambda$-type there-level system. We calculate the fidelity and the absorption of the system in the two modulation situations, and the obtained analytical results agree well with the numerical ones. Our results for the modulated EIT rule out the ATS interpretation and show that the QDI interpretation is more appropriate for our modulated experiments. By including the detuning and the population decay, which present in all kinds of EIT or "EIT-like" experiments, we numerically confirm the QDI's prediction for the modulated EIT under real experimental conditions. Our proposal for the modulated EIT experiments provides an unambiguous way to discern the QDI and the ATS interpretations, and is readily implemented in atomic gases, artificial atoms in superconducting quantum circuits, and three-level meta-atoms in meta-materials.

\begin{acknowledgments}
This work is supported by the National Natural Science Foundation of China under Grant No. 11574239, 11547310, and 11275139, the National Basic Research Program of China Grant No. 2013CB922003, and the Fundamental Research Funds for the Central Universities.
\end{acknowledgments}

\appendix

\section{Master equation with a Lindblad form}
\label{app:lme}
In this Appendix, we evaluate the absorption in Sec.~\ref{sec:dec} and Sec.~\ref{sec:mix} with the Lindblad master equation~\cite{RevModPhys.77.633}, instead of Eq.~(\ref{eq:me}),
\begin{eqnarray}\label{eq:lme}
  \frac{\partial \rho (t)}{\partial t}& = &i[\rho,H_{II}] +\frac{\Gamma_{ab}}{2}\left[2\sigma_{ba}\rho\sigma_{ab}-
  \sigma_{aa}\rho-\rho\sigma_{aa}\right]\nonumber\\
  &&+\frac{\Gamma_{ac}}{2}\left[2\sigma_{ca}\rho\sigma_{ac}-
  \sigma_{aa}\rho-\rho\sigma_{aa}\right]
\end{eqnarray}
with $\sigma_{ij}$ is the atomic projection operator ($i,j=a,b,c$), and the second and third terms on the right-hand side
describe spontaneous emission from the state$|a\rangle$ to the states $|b\rangle$
and $|c\rangle$ , with rates $\Gamma_{ab}$ and $\Gamma_{ac}$, respectively.

\begin{figure}
\includegraphics[width=3in]{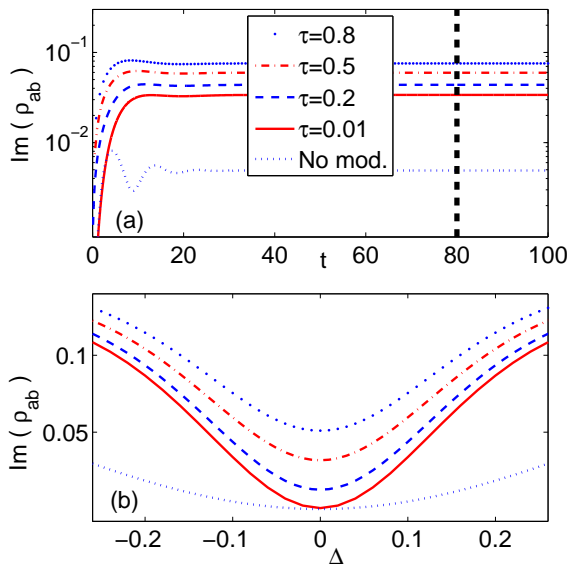}
\caption{\label{fig:atd} (Color online)  Same as Fig.~\ref{fig:td} except employing the Lindblad master equation with $\Gamma_{ab}=\Gamma_{ac}=0.5$. The behaviors are the same as those in Fig.~\ref{fig:td}.}
\end{figure}

\begin{figure}
\includegraphics[width=3in]{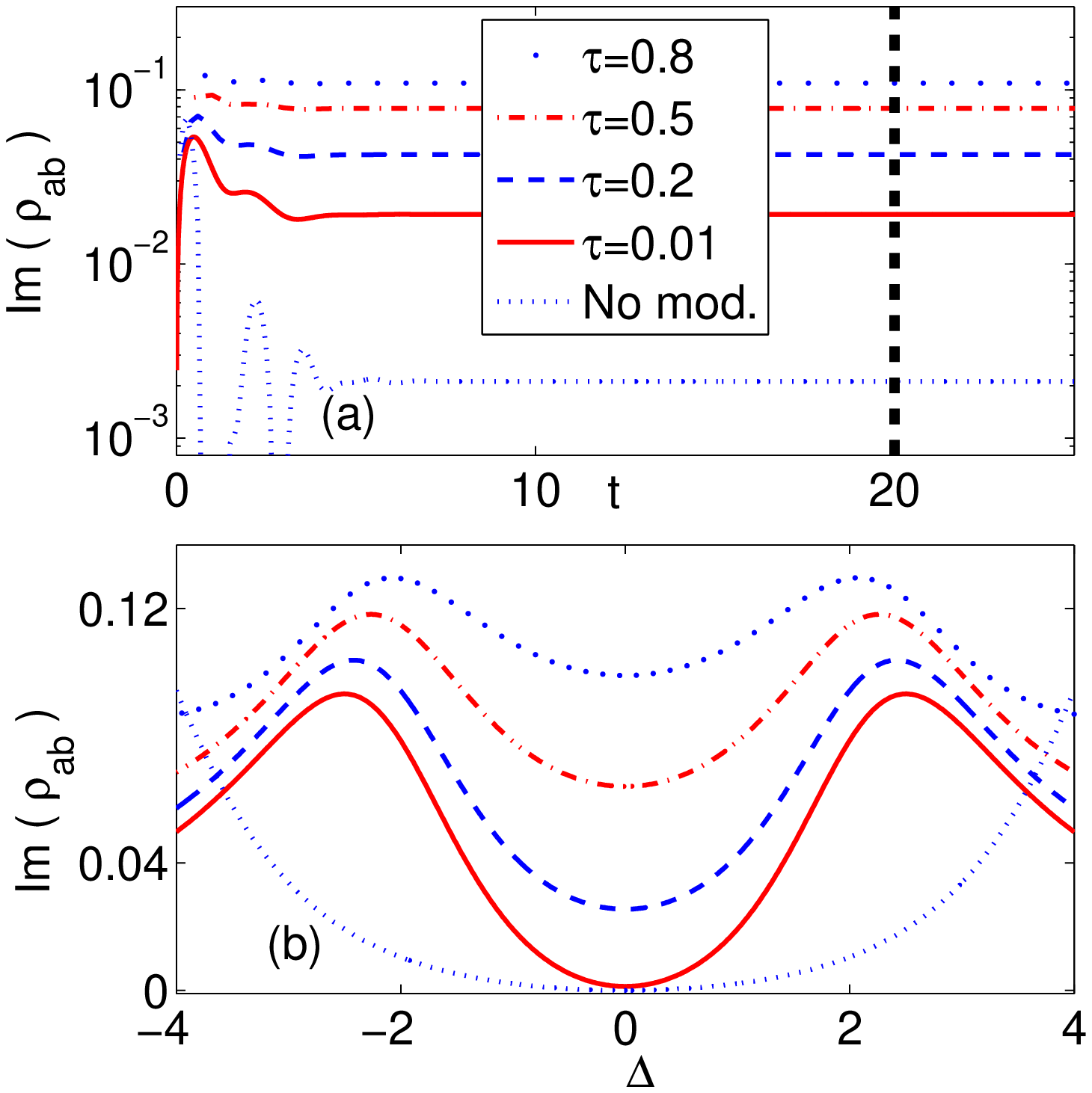}
\caption{\label{fig:ameit} (Color online) Same as Fig.~\ref{fig:meit} except employing the Lindblad master equation with $\Gamma_{ab}=\Gamma_{ac}=2.5$. The behaviors are the same as those in Fig.~\ref{fig:meit}.}
\end{figure}

Figure~\ref{fig:atd} (Fig.~\ref{fig:ameit}) essentially shows the same phenomena as Fig.~\ref{fig:td} (Fig.~\ref{fig:meit}). Therefore adopting the Lindblad master equation does not alter our conclusions.

\begin{thebibliography}{44}
\expandafter\ifx\csname natexlab\endcsname\relax\def\natexlab#1{#1}\fi
\expandafter\ifx\csname bibnamefont\endcsname\relax
  \def\bibnamefont#1{#1}\fi
\expandafter\ifx\csname bibfnamefont\endcsname\relax
  \def\bibfnamefont#1{#1}\fi
\expandafter\ifx\csname citenamefont\endcsname\relax
  \def\citenamefont#1{#1}\fi
\expandafter\ifx\csname url\endcsname\relax
  \def\url#1{\texttt{#1}}\fi
\expandafter\ifx\csname urlprefix\endcsname\relax\def\urlprefix{URL }\fi
\providecommand{\bibinfo}[2]{#2}
\providecommand{\eprint}[2][]{\url{#2}}

\bibitem[{\citenamefont{Boller et~al.}(1991)\citenamefont{Boller,
  Imamo\ifmmode~\breve{g}\else \u{g}\fi{}lu, and Harris}}]{PhysRevLett.66.2593}
\bibinfo{author}{\bibfnamefont{K.-J.} \bibnamefont{Boller}},
  \bibinfo{author}{\bibfnamefont{A.}~\bibnamefont{Imamo\ifmmode~\breve{g}\else
  \u{g}\fi{}lu}}, \bibnamefont{and} \bibinfo{author}{\bibfnamefont{S.~E.}
  \bibnamefont{Harris}}, \bibinfo{journal}{Phys. Rev. Lett.}
  \textbf{\bibinfo{volume}{66}}, \bibinfo{pages}{2593} (\bibinfo{year}{1991}).

\bibitem[{\citenamefont{Qu et~al.}(2013)\citenamefont{Qu, Zhang, and
  Gu}}]{ISI:000325006700104}
\bibinfo{author}{\bibfnamefont{S.~P.} \bibnamefont{Qu}},
  \bibinfo{author}{\bibfnamefont{Y.}~\bibnamefont{Zhang}}, \bibnamefont{and}
  \bibinfo{author}{\bibfnamefont{S.~H.} \bibnamefont{Gu}},
  \bibinfo{journal}{Chin. Phys. B} \textbf{\bibinfo{volume}{22}}
  (\bibinfo{year}{2013}).

\bibitem[{\citenamefont{Xiao et~al.}(1995)\citenamefont{Xiao, Li, Jin, and
  Gea-Banacloche}}]{PhysRevLett.74.666}
\bibinfo{author}{\bibfnamefont{M.}~\bibnamefont{Xiao}},
  \bibinfo{author}{\bibfnamefont{Y.-Q.} \bibnamefont{Li}},
  \bibinfo{author}{\bibfnamefont{S.-Z.} \bibnamefont{Jin}}, \bibnamefont{and}
  \bibinfo{author}{\bibfnamefont{J.}~\bibnamefont{Gea-Banacloche}},
  \bibinfo{journal}{Phys. Rev. Lett.} \textbf{\bibinfo{volume}{74}},
  \bibinfo{pages}{666} (\bibinfo{year}{1995}).

\bibitem[{\citenamefont{Kelly et~al.}(2010)\citenamefont{Kelly, Dutton,
  Schlafer, Mookerji, Ohki, Kline, and Pappas}}]{PhysRevLett.104.163601}
\bibinfo{author}{\bibfnamefont{W.~R.} \bibnamefont{Kelly}},
  \bibinfo{author}{\bibfnamefont{Z.}~\bibnamefont{Dutton}},
  \bibinfo{author}{\bibfnamefont{J.}~\bibnamefont{Schlafer}},
  \bibinfo{author}{\bibfnamefont{B.}~\bibnamefont{Mookerji}},
  \bibinfo{author}{\bibfnamefont{T.~A.} \bibnamefont{Ohki}},
  \bibinfo{author}{\bibfnamefont{J.~S.} \bibnamefont{Kline}}, \bibnamefont{and}
  \bibinfo{author}{\bibfnamefont{D.~P.} \bibnamefont{Pappas}},
  \bibinfo{journal}{Phys. Rev. Lett.} \textbf{\bibinfo{volume}{104}},
  \bibinfo{pages}{163601} (\bibinfo{year}{2010}).

\bibitem[{\citenamefont{Xu et~al.}(2008)\citenamefont{Xu, Sun, Berman, Steel,
  Bracker, Gammon, and Sham}}]{ISI:000259686400011}
\bibinfo{author}{\bibfnamefont{X.}~\bibnamefont{Xu}},
  \bibinfo{author}{\bibfnamefont{B.}~\bibnamefont{Sun}},
  \bibinfo{author}{\bibfnamefont{P.~R.} \bibnamefont{Berman}},
  \bibinfo{author}{\bibfnamefont{D.~G.} \bibnamefont{Steel}},
  \bibinfo{author}{\bibfnamefont{A.~S.} \bibnamefont{Bracker}},
  \bibinfo{author}{\bibfnamefont{D.}~\bibnamefont{Gammon}}, \bibnamefont{and}
  \bibinfo{author}{\bibfnamefont{L.~J.} \bibnamefont{Sham}},
  \bibinfo{journal}{Nature Phys.} \textbf{\bibinfo{volume}{4}},
  \bibinfo{pages}{692} (\bibinfo{year}{2008}).

\bibitem[{\citenamefont{Safavi-Naeini et~al.}(2011)\citenamefont{Safavi-Naeini,
  Alegre, Chan, Eichenfield, Winger, Lin, Hill, Chang, and
  Painter}}]{ISI:000289199400038}
\bibinfo{author}{\bibfnamefont{A.~H.} \bibnamefont{Safavi-Naeini}},
  \bibinfo{author}{\bibfnamefont{T.~P.~M.} \bibnamefont{Alegre}},
  \bibinfo{author}{\bibfnamefont{J.}~\bibnamefont{Chan}},
  \bibinfo{author}{\bibfnamefont{M.}~\bibnamefont{Eichenfield}},
  \bibinfo{author}{\bibfnamefont{M.}~\bibnamefont{Winger}},
  \bibinfo{author}{\bibfnamefont{Q.}~\bibnamefont{Lin}},
  \bibinfo{author}{\bibfnamefont{J.~T.} \bibnamefont{Hill}},
  \bibinfo{author}{\bibfnamefont{D.~E.} \bibnamefont{Chang}}, \bibnamefont{and}
  \bibinfo{author}{\bibfnamefont{O.}~\bibnamefont{Painter}},
  \bibinfo{journal}{Nature(London)} \textbf{\bibinfo{volume}{472}},
  \bibinfo{pages}{69} (\bibinfo{year}{2011}).

\bibitem[{\citenamefont{Zhang et~al.}(2008{\natexlab{a}})\citenamefont{Zhang,
  Genov, Wang, Liu, and Zhang}}]{PhysRevLett.101.047401}
\bibinfo{author}{\bibfnamefont{S.}~\bibnamefont{Zhang}},
  \bibinfo{author}{\bibfnamefont{D.~A.} \bibnamefont{Genov}},
  \bibinfo{author}{\bibfnamefont{Y.}~\bibnamefont{Wang}},
  \bibinfo{author}{\bibfnamefont{M.}~\bibnamefont{Liu}}, \bibnamefont{and}
  \bibinfo{author}{\bibfnamefont{X.}~\bibnamefont{Zhang}},
  \bibinfo{journal}{Phys. Rev. Lett.} \textbf{\bibinfo{volume}{101}},
  \bibinfo{pages}{047401} (\bibinfo{year}{2008}{\natexlab{a}}).

\bibitem[{\citenamefont{Papasimakis et~al.}(2008)\citenamefont{Papasimakis,
  Fedotov, Zheludev, and Prosvirnin}}]{PhysRevLett.101.253903}
\bibinfo{author}{\bibfnamefont{N.}~\bibnamefont{Papasimakis}},
  \bibinfo{author}{\bibfnamefont{V.~A.} \bibnamefont{Fedotov}},
  \bibinfo{author}{\bibfnamefont{N.~I.} \bibnamefont{Zheludev}},
  \bibnamefont{and} \bibinfo{author}{\bibfnamefont{S.~L.}
  \bibnamefont{Prosvirnin}}, \bibinfo{journal}{Phys. Rev. Lett.}
  \textbf{\bibinfo{volume}{101}}, \bibinfo{pages}{253903}
  (\bibinfo{year}{2008}).

\bibitem[{\citenamefont{Phillips et~al.}(2001)\citenamefont{Phillips,
  Fleischhauer, Mair, Walsworth, and Lukin}}]{PhysRevLett.86.783}
\bibinfo{author}{\bibfnamefont{D.~F.} \bibnamefont{Phillips}},
  \bibinfo{author}{\bibfnamefont{A.}~\bibnamefont{Fleischhauer}},
  \bibinfo{author}{\bibfnamefont{A.}~\bibnamefont{Mair}},
  \bibinfo{author}{\bibfnamefont{R.~L.} \bibnamefont{Walsworth}},
  \bibnamefont{and} \bibinfo{author}{\bibfnamefont{M.~D.} \bibnamefont{Lukin}},
  \bibinfo{journal}{Phys. Rev. Lett.} \textbf{\bibinfo{volume}{86}},
  \bibinfo{pages}{783} (\bibinfo{year}{2001}).

\bibitem[{\citenamefont{Novikova et~al.}(2012)\citenamefont{Novikova,
  Walsworth, and Xiao}}]{ISI:000303597400005}
\bibinfo{author}{\bibfnamefont{I.}~\bibnamefont{Novikova}},
  \bibinfo{author}{\bibfnamefont{R.~L.} \bibnamefont{Walsworth}},
  \bibnamefont{and} \bibinfo{author}{\bibfnamefont{Y.}~\bibnamefont{Xiao}},
  \bibinfo{journal}{Laser Photonics Rev.} \textbf{\bibinfo{volume}{6}},
  \bibinfo{pages}{333} (\bibinfo{year}{2012}).

\bibitem[{\citenamefont{Harris et~al.}(1990)\citenamefont{Harris, Field, and
  Imamo\ifmmode~\breve{g}\else \u{g}\fi{}lu}}]{PhysRevLett.64.1107}
\bibinfo{author}{\bibfnamefont{S.~E.} \bibnamefont{Harris}},
  \bibinfo{author}{\bibfnamefont{J.~E.} \bibnamefont{Field}}, \bibnamefont{and}
  \bibinfo{author}{\bibfnamefont{A.}~\bibnamefont{Imamo\ifmmode~\breve{g}\else
  \u{g}\fi{}lu}}, \bibinfo{journal}{Phys. Rev. Lett.}
  \textbf{\bibinfo{volume}{64}}, \bibinfo{pages}{1107} (\bibinfo{year}{1990}).

\bibitem[{\citenamefont{He et~al.}(2007)\citenamefont{He, Liu, Yi, Sun, and
  Nori}}]{PhysRevA.75.063818}
\bibinfo{author}{\bibfnamefont{L.}~\bibnamefont{He}},
  \bibinfo{author}{\bibfnamefont{Y.-X.} \bibnamefont{Liu}},
  \bibinfo{author}{\bibfnamefont{S.}~\bibnamefont{Yi}},
  \bibinfo{author}{\bibfnamefont{C.~P.} \bibnamefont{Sun}}, \bibnamefont{and}
  \bibinfo{author}{\bibfnamefont{F.}~\bibnamefont{Nori}},
  \bibinfo{journal}{Phys. Rev. A} \textbf{\bibinfo{volume}{75}},
  \bibinfo{pages}{063818} (\bibinfo{year}{2007}).

\bibitem[{\citenamefont{Liao et~al.}(2009)\citenamefont{Liao, Huang, Liu,
  Kuang, and Sun}}]{PhysRevA.80.014301}
\bibinfo{author}{\bibfnamefont{J.-Q.} \bibnamefont{Liao}},
  \bibinfo{author}{\bibfnamefont{J.-F.} \bibnamefont{Huang}},
  \bibinfo{author}{\bibfnamefont{Y.-X.} \bibnamefont{Liu}},
  \bibinfo{author}{\bibfnamefont{L.-M.} \bibnamefont{Kuang}}, \bibnamefont{and}
  \bibinfo{author}{\bibfnamefont{C.~P.} \bibnamefont{Sun}},
  \bibinfo{journal}{Phys. Rev. A} \textbf{\bibinfo{volume}{80}},
  \bibinfo{pages}{014301} (\bibinfo{year}{2009}).

\bibitem[{\citenamefont{Yang et~al.}(2015)\citenamefont{Yang, Bao, and
  Pan}}]{PhysRevA.91.053805}
\bibinfo{author}{\bibfnamefont{S.-J.} \bibnamefont{Yang}},
  \bibinfo{author}{\bibfnamefont{X.-H.} \bibnamefont{Bao}}, \bibnamefont{and}
  \bibinfo{author}{\bibfnamefont{J.-W.} \bibnamefont{Pan}},
  \bibinfo{journal}{Phys. Rev. A} \textbf{\bibinfo{volume}{91}},
  \bibinfo{pages}{053805} (\bibinfo{year}{2015}).

\bibitem[{\citenamefont{Liu et~al.}(2005)\citenamefont{Liu, You, Wei, Sun, and
  Nori}}]{PhysRevLett.95.087001}
\bibinfo{author}{\bibfnamefont{Y.-X.} \bibnamefont{Liu}},
  \bibinfo{author}{\bibfnamefont{J.~Q.} \bibnamefont{You}},
  \bibinfo{author}{\bibfnamefont{L.~F.} \bibnamefont{Wei}},
  \bibinfo{author}{\bibfnamefont{C.~P.} \bibnamefont{Sun}}, \bibnamefont{and}
  \bibinfo{author}{\bibfnamefont{F.}~\bibnamefont{Nori}},
  \bibinfo{journal}{Phys. Rev. Lett.} \textbf{\bibinfo{volume}{95}},
  \bibinfo{pages}{087001} (\bibinfo{year}{2005}).

\bibitem[{\citenamefont{Fano}(1961)}]{PhysRev.124.1866}
\bibinfo{author}{\bibfnamefont{U.}~\bibnamefont{Fano}}, \bibinfo{journal}{Phys.
  Rev.} \textbf{\bibinfo{volume}{124}}, \bibinfo{pages}{1866}
  (\bibinfo{year}{1961}).

\bibitem[{\citenamefont{Marangos}(1998)}]{doi:10.1080/09500349808231909}
\bibinfo{author}{\bibfnamefont{J.~P.} \bibnamefont{Marangos}},
  \bibinfo{journal}{J. Mod. Opt.} \textbf{\bibinfo{volume}{45}},
  \bibinfo{pages}{471} (\bibinfo{year}{1998}).

\bibitem[{\citenamefont{Fleischhauer et~al.}(2005)\citenamefont{Fleischhauer,
  Imamoglu, and Marangos}}]{RevModPhys.77.633}
\bibinfo{author}{\bibfnamefont{M.}~\bibnamefont{Fleischhauer}},
  \bibinfo{author}{\bibfnamefont{A.}~\bibnamefont{Imamoglu}}, \bibnamefont{and}
  \bibinfo{author}{\bibfnamefont{J.~P.} \bibnamefont{Marangos}},
  \bibinfo{journal}{Rev. Mod. Phys.} \textbf{\bibinfo{volume}{77}},
  \bibinfo{pages}{633} (\bibinfo{year}{2005}).

\bibitem[{\citenamefont{Autler and Townes}(1955)}]{PhysRev.100.703}
\bibinfo{author}{\bibfnamefont{S.~H.} \bibnamefont{Autler}} \bibnamefont{and}
  \bibinfo{author}{\bibfnamefont{C.~H.} \bibnamefont{Townes}},
  \bibinfo{journal}{Phys. Rev.} \textbf{\bibinfo{volume}{100}},
  \bibinfo{pages}{703} (\bibinfo{year}{1955}).

\bibitem[{\citenamefont{Drell}(1996)}]{ALB}
\bibinfo{author}{\bibfnamefont{S.~D.} \bibnamefont{Drell}},
  \emph{\bibinfo{title}{Amazing Light}} (\bibinfo{publisher}{Springer, New
  York}, \bibinfo{year}{1996}).

\bibitem[{\citenamefont{Davuluri et~al.}(2015)\citenamefont{Davuluri, Wang, and
  Zhu}}]{ISI:000357418300009}
\bibinfo{author}{\bibfnamefont{S.}~\bibnamefont{Davuluri}},
  \bibinfo{author}{\bibfnamefont{Y.}~\bibnamefont{Wang}}, \bibnamefont{and}
  \bibinfo{author}{\bibfnamefont{S.}~\bibnamefont{Zhu}}, \bibinfo{journal}{J.
  Mod. Opt.} \textbf{\bibinfo{volume}{62}}, \bibinfo{pages}{1091}
  (\bibinfo{year}{2015}).

\bibitem[{\citenamefont{Cohen-Tannoudji and
  Reynaud}(1977)}]{0022-3700-10-12-010}
\bibinfo{author}{\bibfnamefont{C.}~\bibnamefont{Cohen-Tannoudji}}
  \bibnamefont{and} \bibinfo{author}{\bibfnamefont{S.}~\bibnamefont{Reynaud}},
  \bibinfo{journal}{J. Phys. B At. Mol. Phys} \textbf{\bibinfo{volume}{10}},
  \bibinfo{pages}{2311} (\bibinfo{year}{1977}).

\bibitem[{\citenamefont{Zhang et~al.}(2013)\citenamefont{Zhang, Wang, Chen,
  Bao, Zhang, Zhao, and Jia}}]{PhysRevA.87.033835}
\bibinfo{author}{\bibfnamefont{H.}~\bibnamefont{Zhang}},
  \bibinfo{author}{\bibfnamefont{L.}~\bibnamefont{Wang}},
  \bibinfo{author}{\bibfnamefont{J.}~\bibnamefont{Chen}},
  \bibinfo{author}{\bibfnamefont{S.}~\bibnamefont{Bao}},
  \bibinfo{author}{\bibfnamefont{L.}~\bibnamefont{Zhang}},
  \bibinfo{author}{\bibfnamefont{J.}~\bibnamefont{Zhao}}, \bibnamefont{and}
  \bibinfo{author}{\bibfnamefont{S.}~\bibnamefont{Jia}},
  \bibinfo{journal}{Phys. Rev. A} \textbf{\bibinfo{volume}{87}},
  \bibinfo{pages}{033835} (\bibinfo{year}{2013}).

\bibitem[{\citenamefont{Harris}(1997)}]{:/content/aip/magazine/physicstoday/article/50/7/10.1063/1.881806}
\bibinfo{author}{\bibfnamefont{S.~E.} \bibnamefont{Harris}},
  \bibinfo{journal}{Phys. Today} \textbf{\bibinfo{volume}{50}}
  (\bibinfo{year}{1997}).

\bibitem[{\citenamefont{Anisimov et~al.}(2011)\citenamefont{Anisimov, Dowling,
  and Sanders}}]{PhysRevLett.107.163604}
\bibinfo{author}{\bibfnamefont{P.~M.} \bibnamefont{Anisimov}},
  \bibinfo{author}{\bibfnamefont{J.~P.} \bibnamefont{Dowling}},
  \bibnamefont{and} \bibinfo{author}{\bibfnamefont{B.~C.}
  \bibnamefont{Sanders}}, \bibinfo{journal}{Phys. Rev. Lett.}
  \textbf{\bibinfo{volume}{107}}, \bibinfo{pages}{163604}
  (\bibinfo{year}{2011}).

\bibitem[{\citenamefont{Giner et~al.}(2013)\citenamefont{Giner, Veissier,
  Sparkes, Sheremet, Nicolas, Mishina, Scherman, Burks, Shomroni, Kupriyanov
  et~al.}}]{PhysRevA.87.013823}
\bibinfo{author}{\bibfnamefont{L.}~\bibnamefont{Giner}},
  \bibinfo{author}{\bibfnamefont{L.}~\bibnamefont{Veissier}},
  \bibinfo{author}{\bibfnamefont{B.}~\bibnamefont{Sparkes}},
  \bibinfo{author}{\bibfnamefont{A.~S.} \bibnamefont{Sheremet}},
  \bibinfo{author}{\bibfnamefont{A.}~\bibnamefont{Nicolas}},
  \bibinfo{author}{\bibfnamefont{O.~S.} \bibnamefont{Mishina}},
  \bibinfo{author}{\bibfnamefont{M.}~\bibnamefont{Scherman}},
  \bibinfo{author}{\bibfnamefont{S.}~\bibnamefont{Burks}},
  \bibinfo{author}{\bibfnamefont{I.}~\bibnamefont{Shomroni}},
  \bibinfo{author}{\bibfnamefont{D.~V.} \bibnamefont{Kupriyanov}},
  \bibnamefont{et~al.}, \bibinfo{journal}{Phys. Rev. A}
  \textbf{\bibinfo{volume}{87}}, \bibinfo{pages}{013823}
  (\bibinfo{year}{2013}).

\bibitem[{\citenamefont{Abi-Salloum}(2010)}]{PhysRevA.81.053836}
\bibinfo{author}{\bibfnamefont{T.~Y.} \bibnamefont{Abi-Salloum}},
  \bibinfo{journal}{Phys. Rev. A} \textbf{\bibinfo{volume}{81}},
  \bibinfo{pages}{053836} (\bibinfo{year}{2010}).

\bibitem[{\citenamefont{Zhu et~al.}(2013)\citenamefont{Zhu, Tan, and
  Huang}}]{PhysRevA.87.043813}
\bibinfo{author}{\bibfnamefont{C.}~\bibnamefont{Zhu}},
  \bibinfo{author}{\bibfnamefont{C.}~\bibnamefont{Tan}}, \bibnamefont{and}
  \bibinfo{author}{\bibfnamefont{G.}~\bibnamefont{Huang}},
  \bibinfo{journal}{Phys. Rev. A} \textbf{\bibinfo{volume}{87}},
  \bibinfo{pages}{043813} (\bibinfo{year}{2013}).

\bibitem[{\citenamefont{Sun et~al.}(2014)\citenamefont{Sun, Liu, Ian, You,
  Il'ichev, and Nori}}]{PhysRevA.89.063822}
\bibinfo{author}{\bibfnamefont{H.-C.} \bibnamefont{Sun}},
  \bibinfo{author}{\bibfnamefont{Y.-X.} \bibnamefont{Liu}},
  \bibinfo{author}{\bibfnamefont{H.}~\bibnamefont{Ian}},
  \bibinfo{author}{\bibfnamefont{J.~Q.} \bibnamefont{You}},
  \bibinfo{author}{\bibfnamefont{E.}~\bibnamefont{Il'ichev}}, \bibnamefont{and}
  \bibinfo{author}{\bibfnamefont{F.}~\bibnamefont{Nori}},
  \bibinfo{journal}{Phys. Rev. A} \textbf{\bibinfo{volume}{89}},
  \bibinfo{pages}{063822} (\bibinfo{year}{2014}).

\bibitem[{\citenamefont{Greentree et~al.}(2002)\citenamefont{Greentree, Smith,
  de~Echaniz, Durrant, Marangos, Segal, and Vaccaro}}]{PhysRevA.65.053802}
\bibinfo{author}{\bibfnamefont{A.~D.} \bibnamefont{Greentree}},
  \bibinfo{author}{\bibfnamefont{T.~B.} \bibnamefont{Smith}},
  \bibinfo{author}{\bibfnamefont{S.~R.} \bibnamefont{de~Echaniz}},
  \bibinfo{author}{\bibfnamefont{A.~V.} \bibnamefont{Durrant}},
  \bibinfo{author}{\bibfnamefont{J.~P.} \bibnamefont{Marangos}},
  \bibinfo{author}{\bibfnamefont{D.~M.} \bibnamefont{Segal}}, \bibnamefont{and}
  \bibinfo{author}{\bibfnamefont{J.~A.} \bibnamefont{Vaccaro}},
  \bibinfo{journal}{Phys. Rev. A} \textbf{\bibinfo{volume}{65}},
  \bibinfo{pages}{053802} (\bibinfo{year}{2002}).

\bibitem[{\citenamefont{Wu and Yang}(2007)}]{PhysRevLett.98.013601}
\bibinfo{author}{\bibfnamefont{Y.}~\bibnamefont{Wu}} \bibnamefont{and}
  \bibinfo{author}{\bibfnamefont{X.}~\bibnamefont{Yang}},
  \bibinfo{journal}{Phys. Rev. Lett.} \textbf{\bibinfo{volume}{98}},
  \bibinfo{pages}{013601} (\bibinfo{year}{2007}).

\bibitem[{\citenamefont{Jing et~al.}(2015)\citenamefont{Jing, Wu, Byrd, You,
  Yu, and Wang}}]{PhysRevLett.114.190502}
\bibinfo{author}{\bibfnamefont{J.}~\bibnamefont{Jing}},
  \bibinfo{author}{\bibfnamefont{L.-A.} \bibnamefont{Wu}},
  \bibinfo{author}{\bibfnamefont{M.}~\bibnamefont{Byrd}},
  \bibinfo{author}{\bibfnamefont{J.~Q.} \bibnamefont{You}},
  \bibinfo{author}{\bibfnamefont{T.}~\bibnamefont{Yu}}, \bibnamefont{and}
  \bibinfo{author}{\bibfnamefont{Z.-M.} \bibnamefont{Wang}},
  \bibinfo{journal}{Phys. Rev. Lett.} \textbf{\bibinfo{volume}{114}},
  \bibinfo{pages}{190502} (\bibinfo{year}{2015}).

\bibitem[{\citenamefont{Zhang et~al.}(2007)\citenamefont{Zhang, Dobrovitski,
  Santos, Viola, and Harmon}}]{PhysRevB.75.201302}
\bibinfo{author}{\bibfnamefont{W.}~\bibnamefont{Zhang}},
  \bibinfo{author}{\bibfnamefont{V.~V.} \bibnamefont{Dobrovitski}},
  \bibinfo{author}{\bibfnamefont{L.~F.} \bibnamefont{Santos}},
  \bibinfo{author}{\bibfnamefont{L.}~\bibnamefont{Viola}}, \bibnamefont{and}
  \bibinfo{author}{\bibfnamefont{B.~N.} \bibnamefont{Harmon}},
  \bibinfo{journal}{Phys. Rev. B} \textbf{\bibinfo{volume}{75}},
  \bibinfo{pages}{201302} (\bibinfo{year}{2007}).

\bibitem[{\citenamefont{Zhang et~al.}(2008{\natexlab{b}})\citenamefont{Zhang,
  Konstantinidis, Dobrovitski, Harmon, Santos, and Viola}}]{PhysRevB.77.125336}
\bibinfo{author}{\bibfnamefont{W.}~\bibnamefont{Zhang}},
  \bibinfo{author}{\bibfnamefont{N.~P.} \bibnamefont{Konstantinidis}},
  \bibinfo{author}{\bibfnamefont{V.~V.} \bibnamefont{Dobrovitski}},
  \bibinfo{author}{\bibfnamefont{B.~N.} \bibnamefont{Harmon}},
  \bibinfo{author}{\bibfnamefont{L.~F.} \bibnamefont{Santos}},
  \bibnamefont{and} \bibinfo{author}{\bibfnamefont{L.}~\bibnamefont{Viola}},
  \bibinfo{journal}{Phys. Rev. B} \textbf{\bibinfo{volume}{77}},
  \bibinfo{pages}{125336} (\bibinfo{year}{2008}{\natexlab{b}}).

\bibitem[{\citenamefont{Li and Xiao}(1995)}]{PhysRevA.51.4959}
\bibinfo{author}{\bibfnamefont{Y.-Q.} \bibnamefont{Li}} \bibnamefont{and}
  \bibinfo{author}{\bibfnamefont{M.}~\bibnamefont{Xiao}},
  \bibinfo{journal}{Phys. Rev. A} \textbf{\bibinfo{volume}{51}},
  \bibinfo{pages}{4959} (\bibinfo{year}{1995}).

\bibitem[{\citenamefont{Scully and Zubairy}(1997)}]{QO}
\bibinfo{author}{\bibfnamefont{M.}~\bibnamefont{Scully}} \bibnamefont{and}
  \bibinfo{author}{\bibfnamefont{M.}~\bibnamefont{Zubairy}},
  \emph{\bibinfo{title}{Quantum Optics}} (\bibinfo{publisher}{Cambridge
  University Press, Cambridge, England}, \bibinfo{year}{1997}).

\bibitem[{\citenamefont{Liu et~al.}(2014)\citenamefont{Liu, Sun, Peng,
  Miranowicz, Tsai, and Nori}}]{ISI:000346274500001}
\bibinfo{author}{\bibfnamefont{Y.-X.} \bibnamefont{Liu}},
  \bibinfo{author}{\bibfnamefont{H.-C.} \bibnamefont{Sun}},
  \bibinfo{author}{\bibfnamefont{Z.~H.} \bibnamefont{Peng}},
  \bibinfo{author}{\bibfnamefont{A.}~\bibnamefont{Miranowicz}},
  \bibinfo{author}{\bibfnamefont{J.~S.} \bibnamefont{Tsai}}, \bibnamefont{and}
  \bibinfo{author}{\bibfnamefont{F.}~\bibnamefont{Nori}},
  \bibinfo{journal}{Sci. Rep.} \textbf{\bibinfo{volume}{4}}
  (\bibinfo{year}{2014}).

\bibitem[{\citenamefont{Kurter et~al.}(2011)\citenamefont{Kurter, Tassin,
  Zhang, Koschny, Zhuravel, Ustinov, Anlage, and
  Soukoulis}}]{PhysRevLett.107.043901}
\bibinfo{author}{\bibfnamefont{C.}~\bibnamefont{Kurter}},
  \bibinfo{author}{\bibfnamefont{P.}~\bibnamefont{Tassin}},
  \bibinfo{author}{\bibfnamefont{L.}~\bibnamefont{Zhang}},
  \bibinfo{author}{\bibfnamefont{T.}~\bibnamefont{Koschny}},
  \bibinfo{author}{\bibfnamefont{A.~P.} \bibnamefont{Zhuravel}},
  \bibinfo{author}{\bibfnamefont{A.~V.} \bibnamefont{Ustinov}},
  \bibinfo{author}{\bibfnamefont{S.~M.} \bibnamefont{Anlage}},
  \bibnamefont{and} \bibinfo{author}{\bibfnamefont{C.~M.}
  \bibnamefont{Soukoulis}}, \bibinfo{journal}{Phys. Rev. Lett.}
  \textbf{\bibinfo{volume}{107}}, \bibinfo{pages}{043901}
  (\bibinfo{year}{2011}).

\bibitem[{\citenamefont{Zhang et~al.}(2011)\citenamefont{Zhang, Zhang, Yang,
  and Chen}}]{PhysRevE.83.046604}
\bibinfo{author}{\bibfnamefont{L.}~\bibnamefont{Zhang}},
  \bibinfo{author}{\bibfnamefont{Y.}~\bibnamefont{Zhang}},
  \bibinfo{author}{\bibfnamefont{Y.}~\bibnamefont{Yang}}, \bibnamefont{and}
  \bibinfo{author}{\bibfnamefont{H.}~\bibnamefont{Chen}},
  \bibinfo{journal}{Phys. Rev. E} \textbf{\bibinfo{volume}{83}},
  \bibinfo{pages}{046604} (\bibinfo{year}{2011}).

\bibitem[{\citenamefont{Chu et~al.}(2003)\citenamefont{Chu, Liu, and
  Sun}}]{ISI:000182372200013}
\bibinfo{author}{\bibfnamefont{S.}~\bibnamefont{Chu}},
  \bibinfo{author}{\bibfnamefont{T.}~\bibnamefont{Liu}}, \bibnamefont{and}
  \bibinfo{author}{\bibfnamefont{C.}~\bibnamefont{Sun}}, \bibinfo{journal}{Opt.
  Express} \textbf{\bibinfo{volume}{11}}, \bibinfo{pages}{933}
  (\bibinfo{year}{2003}).

\bibitem[{\citenamefont{Van-Brunt and Visser}(2015)}]{ISI:000354870900011}
\bibinfo{author}{\bibfnamefont{A.}~\bibnamefont{Van-Brunt}} \bibnamefont{and}
  \bibinfo{author}{\bibfnamefont{M.}~\bibnamefont{Visser}},
  \bibinfo{journal}{J. Phys. A} \textbf{\bibinfo{volume}{48}}
  (\bibinfo{year}{2015}).

\bibitem[{\citenamefont{Kash et~al.}(1999)\citenamefont{Kash, Sautenkov,
  Zibrov, Hollberg, Welch, Lukin, Rostovtsev, Fry, and
  Scully}}]{PhysRevLett.82.5229}
\bibinfo{author}{\bibfnamefont{M.~M.} \bibnamefont{Kash}},
  \bibinfo{author}{\bibfnamefont{V.~A.} \bibnamefont{Sautenkov}},
  \bibinfo{author}{\bibfnamefont{A.~S.} \bibnamefont{Zibrov}},
  \bibinfo{author}{\bibfnamefont{L.}~\bibnamefont{Hollberg}},
  \bibinfo{author}{\bibfnamefont{G.~R.} \bibnamefont{Welch}},
  \bibinfo{author}{\bibfnamefont{M.~D.} \bibnamefont{Lukin}},
  \bibinfo{author}{\bibfnamefont{Y.}~\bibnamefont{Rostovtsev}},
  \bibinfo{author}{\bibfnamefont{E.~S.} \bibnamefont{Fry}}, \bibnamefont{and}
  \bibinfo{author}{\bibfnamefont{M.~O.} \bibnamefont{Scully}},
  \bibinfo{journal}{Phys. Rev. Lett.} \textbf{\bibinfo{volume}{82}},
  \bibinfo{pages}{5229} (\bibinfo{year}{1999}).

\bibitem[{\citenamefont{Xiao et~al.}(2009)\citenamefont{Xiao, Wang, Baryakhtar,
  Van~Camp, Crescimanno, Hohensee, Jiang, Phillips, Lukin, Yelin
  et~al.}}]{PhysRevA.80.041805}
\bibinfo{author}{\bibfnamefont{Y.}~\bibnamefont{Xiao}},
  \bibinfo{author}{\bibfnamefont{T.}~\bibnamefont{Wang}},
  \bibinfo{author}{\bibfnamefont{M.}~\bibnamefont{Baryakhtar}},
  \bibinfo{author}{\bibfnamefont{M.}~\bibnamefont{Van~Camp}},
  \bibinfo{author}{\bibfnamefont{M.}~\bibnamefont{Crescimanno}},
  \bibinfo{author}{\bibfnamefont{M.}~\bibnamefont{Hohensee}},
  \bibinfo{author}{\bibfnamefont{L.}~\bibnamefont{Jiang}},
  \bibinfo{author}{\bibfnamefont{D.~F.} \bibnamefont{Phillips}},
  \bibinfo{author}{\bibfnamefont{M.~D.} \bibnamefont{Lukin}},
  \bibinfo{author}{\bibfnamefont{S.~F.} \bibnamefont{Yelin}},
  \bibnamefont{et~al.}, \bibinfo{journal}{Phys. Rev. A}
  \textbf{\bibinfo{volume}{80}}, \bibinfo{pages}{041805}
  (\bibinfo{year}{2009}).

\bibitem[{\citenamefont{Xiao et~al.}(2008)\citenamefont{Xiao, Novikova,
  Phillips, and Walsworth}}]{ISI:000259270600066}
\bibinfo{author}{\bibfnamefont{Y.}~\bibnamefont{Xiao}},
  \bibinfo{author}{\bibfnamefont{I.}~\bibnamefont{Novikova}},
  \bibinfo{author}{\bibfnamefont{D.~F.} \bibnamefont{Phillips}},
  \bibnamefont{and} \bibinfo{author}{\bibfnamefont{R.~L.}
  \bibnamefont{Walsworth}}, \bibinfo{journal}{Opt. Express}
  \textbf{\bibinfo{volume}{16}}, \bibinfo{pages}{14128} (\bibinfo{year}{2008}).

\end{thebibliography}

\end{document}